\documentclass{article}
\usepackage{spconf,amssymb,amsmath,epsfig}
\usepackage{textcomp}
\usepackage{amsfonts,amsthm,amsopn}
\usepackage{bm}
\usepackage{hyperref}
\usepackage{url}
\usepackage[capitalise]{cleveref}
\usepackage{graphicx,caption,lipsum}
\usepackage{booktabs,multirow}
\usepackage{placeins}
\usepackage{algpseudocode,algorithm}
\usepackage{color}
\usepackage[table,xcdraw]{xcolor}
\usepackage{bbm}
\usepackage{enumerate}   
\usepackage{adjustbox}
\usepackage[normalem]{ulem}
\usepackage{fixltx2e}
\usepackage[toc,page]{appendix}

\usepackage{makecell}

\ninept
\def\reg{{\rm\ooalign{\hfil
      \raise.07ex\hbox{\scriptsize R}\hfil\crcr\mathhexbox20D}}}

\usepackage{tikz}

\title{Self-supervised text-independent speaker verification using prototypical momentum contrastive learning}

\name{Wei Xia$^{1*}$, Chunlei Zhang$^{2}$, Chao Weng$^{2}$, Meng Yu$^{2}$, Dong Yu$^{2}$ \thanks{* This work was performed when W. Xia was an intern at Tencent AI Lab, Bellevue, WA.
}}
\address{
 $^1$The University of Texas at Dallas, Richardson, TX, USA\\
 $^2$Tencent AI Lab, Bellevue, WA, USA \\
  \small \tt wei.xia@utdallas.edu, \{cleizhang, cweng, raymondmyu, dyu\}@tencent.com}

\newcommand{\vct}[1]{\boldsymbol{\mathbf{#1}}} 

\graphicspath{{./image/}}

\begin{document}

\maketitle

\ninept

\begin{abstract}
  In this study, we investigate self-supervised representation learning for speaker verification (SV). First, we examine a simple contrastive learning approach (SimCLR) with a momentum contrastive (MoCo) learning framework, where the MoCo speaker embedding system utilizes a queue to maintain a large set of negative examples. We show that better speaker embeddings can be learned by momentum contrastive learning. Next, alternative augmentation strategies are explored to normalize extrinsic speaker variabilities of two random segments from the same speech utterance. Specifically, augmentation in the waveform largely improves the speaker representations for SV tasks. The proposed MoCo speaker embedding is further improved when a prototypical memory bank is introduced, which encourages the speaker embeddings to be closer to their assigned prototypes with an intermediate clustering step. In addition, we generalize the self-supervised framework to a semi-supervised scenario where only a small portion of the data is labeled. Comprehensive experiments on the Voxceleb dataset demonstrate that our proposed self-supervised approach achieves competitive performance compared with existing techniques, and can approach fully supervised results with partially labeled data.
\end{abstract}
%
\vspace{-2ex}

\section{Introduction}
\label{sec:intro}
\vspace{-1ex}
Speaker verification (SV) is a task to verify a person's identity from audio streams. It provides a natural and efficient way for biometric identity authentication. SV can be directly employed in audio forensics, computer access control, and telephone voice authentication for long-distance calling. The learned speaker representation is also a key component in speaker diarization, text to speech synthesis, voice conversion, automatic speech recognition as well as targeted speech enhancement and separation systems~\cite{ji2020speaker,zhang2020durian,wang2018voicefilter,wang2018speaker}.

State-of-the-art approaches for learning speaker representations usually rely on supervised deep neural networks and large labeled speaker corpora.  Various neural network architectures~\cite{snyder2018x,hajavi2019deep,zhang2017end} were explored as a speaker embedding extractor. Novel loss functions~\cite{liu2017sphereface, deng2019arcface,zhang2018text,chung2020defence} were investigated to learn more discriminative speaker embeddings.
New temporal pooling methods~\cite{okabe2018attentive, xie2019utterance} were presented to aggregate a variable-length audio input to a fixed length utterance level representation.
Noise and language robust speaker recognition models and training paradigms have been proposed and improve SV systems' performance~\cite{yu2017adversarial,xia2019cross}.

Though impressive progress has been achieved with fully supervised models, there remain many challenges for SV systems.
For example, fully supervised models with a limited amount of labels usually do not perform well in a domain mismatched condition~\cite{chung2018voxceleb2, zhang2019utd}.
In many practical cases, it is very costly and difficult to obtain sufficient data annotations. 
Self-supervised learning (SSL) that only requires a large amount of unlabeled data provides an alternative solution for learning representations from speech. 
This study mainly focuses on self-supervised contrastive learning approaches, where the goal is to discriminate positive and negative examples. The popular contrastive learning frameworks, such as SimCLR or MoCo, are usually constructed by an InfoNCE loss, neural network-based encoders and a data augmentation module for providing different views of the same sample~\cite{oord2018representation, chen2020simple, he2020momentum, li2020prototypical}. The InfoNCE loss guides the network training, so the distance of a positive pair in the latent space (i.e., embeddings) is forced to be small, and distances of negative pairs are large. Recently, in~\cite{ding2020learning, inoue2020semi, stafylakis2019self, huh2020augmentation}, researchers explored self-supervised contrastive learning for speaker verification as an unsupervised learning framework or a pre-training method for subsequent tasks. They have shown the effectiveness and the potential of self-supervised contrastive learning-based speaker embeddings.

\vspace{-1ex}
In this work, we focus on effectively utilizing unlabeled and partially labeled speech data in the constrastive learning frameworks for speaker representation learning. First, we investigate and compare SimCLR and MoCo as our contrastive speaker representation learning frameworks. To remove extrinsic speaker variabilities (channel, noise or spectrum distortions) of two random chunks from an utterance, two kinds of augmentation strategies - WavAugment and SpecAugment are investigated. We find that the SpecAug module does not help with our proposed contrastive speaker embedding approaches, while the WavAug module improves the speaker verification performance by a large margin.
Moreover, we introduce a prototypical memory bank to alleviate the class collision problem~\cite{arora2019theoretical}. Since instance discrimination in both SimCLR and MoCo treats every other example as a negative sample, the positive examples may be categorized as negative wrongly. We perform clustering on all samples every epoch and store the prototypes in a memory bank. Negatives prototypes are only sampled from different disjoint classes. Experimental results on the Voxceleb corpus show that with the WavAug module and the prototype memory bank, the Equal Error Rate (EER) of the MoCo based SV system decreases from 15.11\% to 8.23\%, a relative 45.5\% reduction.
Finally, we generalize our proposed self-supervised speaker embedding system to a semi-supervised condition, with partially labeled speakers involved in the training. The results show that our proposed semi-supervised speaker embedding system can approach fully supervised results of the whole labeled set with limited labeled speakers.


\section{Contrastive Speaker representation learning}
\label{sec:model}
\begin{figure}[tbp]
    \centering
    \includegraphics[width=0.9\linewidth, height=4cm]{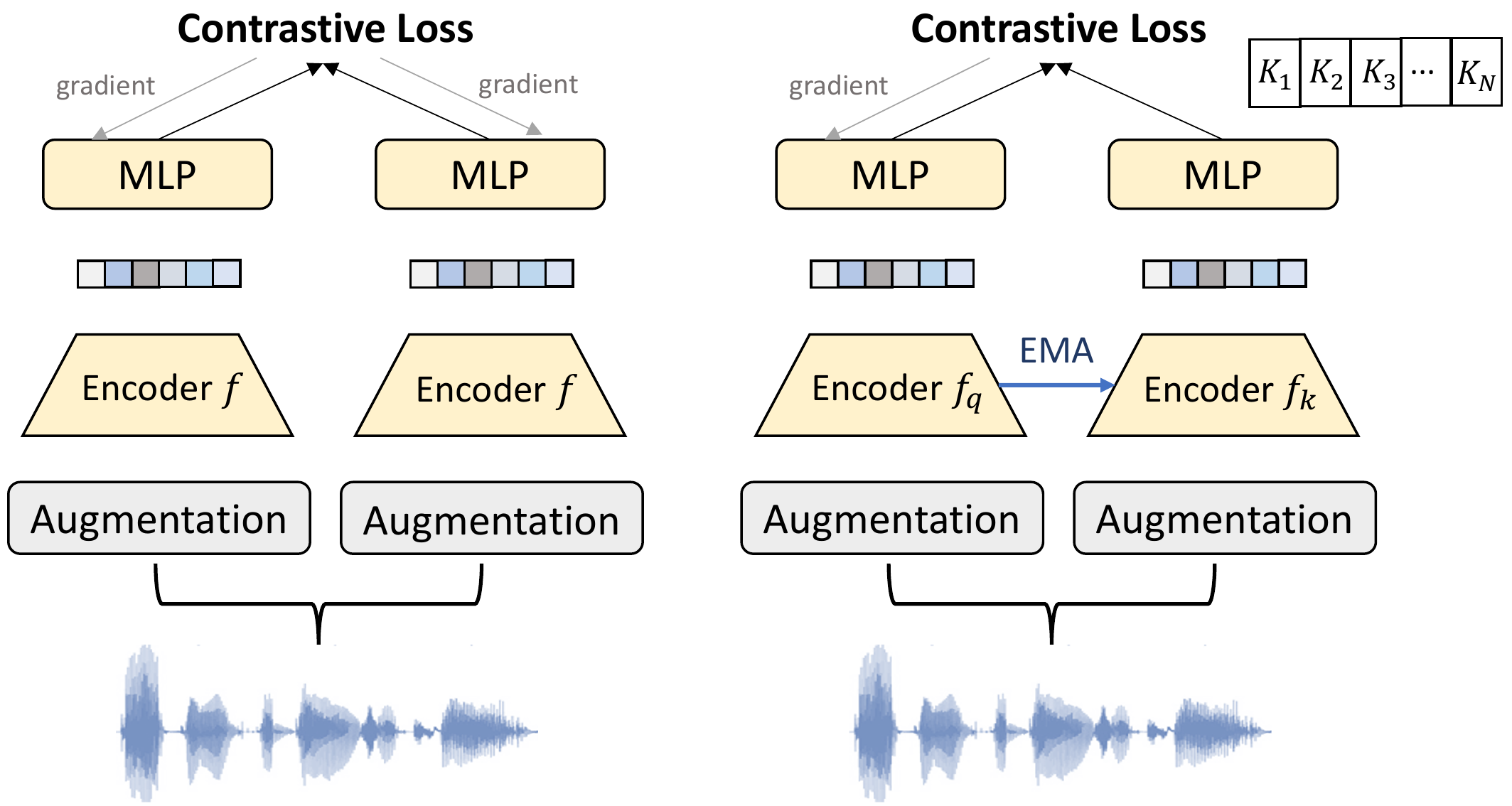}
    \caption{Left: SimCLR approach; Right: Momentum Contrastive learning method for speaker representation learning.}
    \label{fig:se1}
    \vspace{-3ex}
\end{figure}

To learn a good speaker representation, we want the learned embeddings to have the property of inter-class separation and intra-class compactness. This can be achieved by a contrastive learning framework.
It comprises three major components:
(1) A data augmentation module that transforms a data example into two different views, which is a positive pair.
(2) A neural encoder module that encodes the input features into a fixed dimensional embedding.
(3) A projection head that maps embeddings to the latent space where a contrastive loss is applied. 
In the following sections, we investigate two contrastive speaker representation frameworks and introduce a novel prototypical contrastive speaker embedding method.

\vspace{-1ex}
\subsection{SimCLR based speaker embedding}
SimCLR directly maximizes the similarity between augmented positive pairs and minimizes the similarity of negative pairs via a contrastive InfoNCE loss~\cite{oord2018representation} in the latent space.
For a mini-batch of $N$ samples, we can get $2N$ samples after augmentation. SimCLR doesn't sample negative examples explicitly. The $2(N-1)$ samples different from the positive pair are treated as negative examples. The loss function is defined as,
\begin{align}
   \mathcal{L}_{SimCLR}=\frac{1}{N}\sum_{i=1}^{N} -\log \frac{\exp \left( \vct{v}_{i} \cdot \vct{v}^{\prime}_{i} / \tau\right)} 
   {\sum_{j=1}^{2N} \mathbbm{1}_{[i \neq j]} \exp \left( \vct{v}_{i} \cdot \vct{v}_{j} / \tau\right)}
\end{align}
where $ \vct{v}_{i}$ and $\vct{v}^{\prime}_{i}$ are the two augmentations of one sample, $ \vct{v}_{j}$ is from the rest $2N-1$ augmented samples, $\mathbbm{1}_{[i \neq j]} $ is an indicator function that equals to 1 when $i \neq j$. $\tau$ is the temperature parameter. Our above definition is slightly different from the original SimCLR implementation, which also swaps sample $\vct{v}_{i}$ and $\vct{v}^{\prime}_{i}$ and calculates an average loss.   

\vspace{-1ex}
\subsection{Momentum contrastive speaker embedding}
SimCLR uses samples in the current mini-batch for negative example mining.
The negative sample size is constrained with the mini-batch size $N$, which is limited by the GPU memory. To prevent this problem and introduce more negative samples in the InfoNCE loss, we also explore a momentum contrastive based speaker embedding system. We mainly follow the MoCo framework for visual representation learning~\cite{he2020momentum}. MoCo uses a dynamic queue to sample a large number of negative samples to calculate the loss with an assumption that robust features can be learned by a large dictionary that covers a rich set of negative samples. Illustrated in \cref{fig:se1}, one major difference between MoCo and SimCLR is that MoCo uses one query encoder (left) to encode one copy of the augmented samples, and employs a momentum encoder (right) to encode the other copy. So the InfoNCE loss employed in the momentum contrastive speaker embedding learning is defined as,
\vspace{-1ex}
\begin{align}\label{moco_eq}
\mathcal{L}_{MoCo} =\frac{1}{N}\sum_{i=1}^{N}-\log \frac{\exp \left(\vct v_{qi} \cdot \vct v^{\prime}_{{ki}} / \tau\right)}{\sum_{j=0}^{K} \exp \left(\vct v_{qi} \cdot \vct v_{kj} / \tau\right)}
\end{align}

\vspace{-0.5ex}
\noindent 
where $\vct{v}_{qi}$ and $\vct v^{\prime}_{{ki}}$ is one of the query samples (i.e., $anchor$) and corresponding key samples (i.e., $positive$) in the mini-batch. $\vct v_{{kj}} $ are the key samples encoded with the key encoder (momentum encoder), the sum in \cref{moco_eq} is over one $positive$ pair and $K$ $negative$ pairs. The introduction of a queue with the dictionary size $K$ is able to enlarge the number of negative samples compared with the original InfoNCE. At the same time, samples in the dictionary are progressively replaced. $K$ is set to $10000$ in our experiments. When the current mini-batch is enqueued to the dictionary, the oldest mini-batch will be dequeued.

To make the key embeddings in the negative queue consistent, the key encoder $f_k$ is updated as an Exponential Moving Average (EMA) of the query encoder $f_q$. Denoting the parameters of $f_k$ as $\theta_k$ and those of $f_q$ as $\theta_q$, we update $\theta_k$ by $\theta_{\mathrm{k}} \leftarrow m \theta_{\mathrm{k}}+(1-m) \theta_{\mathrm{q}}$, where $m$ = 0.999 is the momentum coefficient. Only the parameters $\theta_q$ are updated by back-propagation during training. 
The momentum update makes the key encoder evolve more smoothly than the query encoder. 
With this design, though the keys in the queue are encoded by different encoders (in different mini-batches), the difference among these encoders is small. Therefore we can maintain a sizeable negative sample queue with a stable training process.

\vspace{-1ex}
\subsection{Prototypical memory bank}
\begin{figure}[tbp]
    \centering
    \includegraphics[width=0.9\linewidth, height=3.8cm]{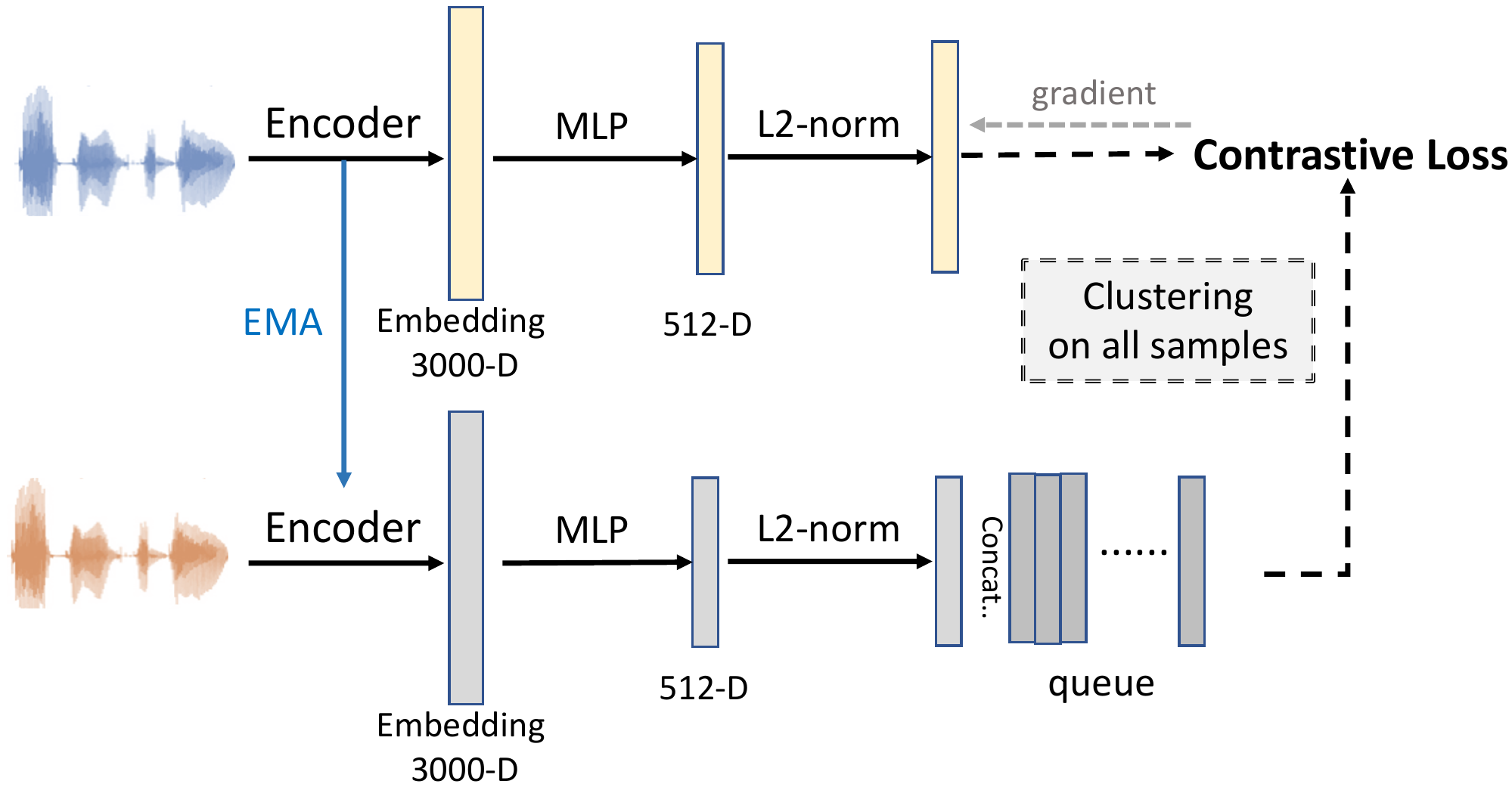}
    \caption{Prototypical momentum contrastive speaker representation learning pipeline.}
    \label{fig:se2}
    \vspace{-3ex}
\end{figure}

MoCo's negative samples in the queue are randomly sampled. It may contain wrong $(anchor, negative)$ pairs since we treat all the other samples as negative. This is known as the class collision problem in self-supervised contrastive learning~\cite{arora2019theoretical}. 
To alleviate this problem, a prototype memory bank is introduced. As shown in \cref{fig:se2}, we perform clustering on all samples every epoch, and negatives prototypes are only sampled from different disjoint classes. Positives are from the same class prototypes. Cluster density is dynamically estimated in order to use a more accurate concentration coefficient ($\tau$ in \cref{moco_eq}) for each cluster.
We can formulate this as an iterative self-labeling process:
\vspace{-0.5ex}
\begin{enumerate}
  \item Train the MoCo with InfoNCE loss for certain epochs in order to make the learned speaker embedding robust and stable for clustering.\vspace{-0.5ex}
  \item  Cluster the whole dataset into $M$ clusters every epoch and store the prototypes (centroids of clusters) in a memory bank.\vspace{-0.5ex}
  \item  Use the pseudo-labels to train a prototypical NCE loss. For an anchor, we pick a positive prototype from the prototype memory bank. From the rest $M-1$ classes, we randomly sample $R$ negative prototypes w.r.t. their  pseudo-labels. 
\end{enumerate}
\noindent 
The overall objective function $\mathcal{L}_{joint}$ is formulated as, \vspace{-0.4ex}
{\scriptsize{
\begin{align}\label{protonce}
-\frac{1}{N}\sum_{i=1}^{N}  \log \frac{\exp \left(\vct{v}_{qi} \cdot \vct{v}^{\prime}_{ki} / \tau\right)}
{\sum_{j=0}^{K} \exp \left(\vct{v}_{qi} \cdot \vct{v}_{kj} / \tau\right)} 
+ \alpha \log \frac{\exp \left(\vct{v}_{qi} \cdot \vct{c}_{s} / \phi_{s}\right)}
{\sum_{j=0}^{R} \exp \left(\vct{v}_{qi} \cdot \vct{c}_{j} / \phi_{j}\right)}
\end{align}}
}

\vspace{-2ex}

\noindent The first term InfoNCE loss in \cref{protonce} is used to retain the property of local smoothness and help bootstrap clustering. 
The second term is denoted as the ProtoNCE loss. $\alpha$ is a weight coefficient to balance two losses. It is set to 0.25 in our experiments. Embedding $\vct{c}_s$ is the prototype of class $s$ and a positive example of the query embedding $\vct{v}_{qi}$. Different class prototypes $\{\vct{c}_j\}_{j=1}^{R}$ are the negative examples. The dynamically estimated temperature coefficient, $\phi=\frac{\sum_{i=1}^{Z}\left\|\vct{v}_{i} - \vct{c} \right\|_{2}}{Z \log (Z+\epsilon)}$, indicates the concentration level of embeddings $\{\vct{v}_i\}_{i=1}^{Z}$ around their class centroid $\vct c$. $Z$ is total number of embeddings in that class and $\epsilon$ is a small constant for numerical stability.

\subsection{Generalization to semi-supervised contrastive learning}
In practice, we hope to approach the fully supervised results with a limited amount of labeled data. The self-supervised contrastive loss is generalized to the semi-supervised condition in the following. First, we explore a generalized supervised contrastive loss (SupCon)~\cite{khosla2020supervised} to utilize arbitrary numbers of positives belonging to the same class. For a sample $i$, the loss $\mathcal{L}_{SupCon}^{i}$ is presented as,
 \small{\begin{align}
\frac{-1}{2 N_{y_{i}}-1} \sum_{p=1}^{2 N} \mathbbm{1}_{[i \neq p,y_{i}=y_{p}]}  \log \frac{\exp \left(\vct{v}_{i} \cdot \vct{v}_{p} / \tau\right)}{\sum_{j=1}^{2 N} \mathbbm{1}_{[i \neq j]}  \exp \left(\vct{v}_{i} \cdot \vct{v}_{j} / \tau\right)}
\end{align}}
\noindent 
For each example, we still generate two augmentations. Since the labels are known, \textit{all} positives $\vct{v}_{p}$ in a mini-batch (the augmented one and any of the remaining samples with the same class label $y_i$) are used in the loss. $N_{y_{i}}$ is the total number of samples with the same label $y_i$ in the mini-batch.
In a partially labeled condition, we add the SupCon loss to the self-supervised MoCo loss and train the model jointly with the following semi-supervised loss, 
\begin{align}
    \mathcal{L}_{semi} = \mathcal{L}^{v}_{SupCon} + \lambda\mathcal{L}^{u+v}_{MoCo}
\end{align}
\noindent where $u$ and $v$ denote the unlabeled and labeled data in a mini-batch, respectively. Parameter $\lambda$ is a weight coefficient to adjust the loss. In a mini-batch, self-supervised MoCo loss is trained on both unlabeled and labeled sets and SupCon loss is trained on the labeled set.


\section{Experimental Setup}
\label{sec:exp}

\subsection{Dataset and feature extraction}
Experiments are performed on the Voxceleb corpus to study the effectiveness of our presented methods. The proposed speaker model is trained only on the VoxCeleb2 \textit{dev} set which contains over 1 million utterances from 5,994 speakers. We evaluate our model on the Voxceleb1 \textit{test} set. There are 40 speakers in the test set with 4874 utterances. The performance is reported in terms of Equal Error Rate (EER) and minimum Detection Cost Function (DCF) with $P_{target} = 0.01$. 

We compute 30-dimensional MFCC on the frame level as input features. A Hamming window of length 25 ms with a 10 ms frame-shift is used to extract the MFCC from input audio signals.
We use a random chunk of 200-400 frame features of each audio file as the input to the network.
The input feature is mean normalized on the frame level. An energy-based VAD is used to remove silent frames.

\subsection{Data augmentation}

In the self-supervised learning frameworks we proposed above, the learned representation is expected to be robust against to the extrinsic variabilities. The data augmentation module is essential as it provides multiple views of the raw signal. In this study, we intend to factor out other information like background noise, interference speakers, or channel variabilities of two random chunks from an utterance as much as possible. Two kinds of augmentation strategies are investigated: WavAug and SpecAug, on the time and frequency domain respectively.

\noindent\textbf{WavAug}. Conventionally in speaker recognition, only one kind of extrinsic variabilities is augmented to the raw sample. In our WavAug module, we find that \textit{progressive} data augmentation is more effective for self-supervised training. Specifically, we first augment two chunks from the same utterance with reverberation by randomly selecting RIRs in the OpenSLR Room Impulse Response and Noise Database. 200 rooms are sampled, and 100 RIRs are picked in each room based on the speaker and receiver position. We set a 80\% probability that the sample to be reverb-augmented, so about 20\% of the whole data is kept unchanged. After the reverb-augmentation, we further add one of the three kinds of noises with equal probability: 1) Noise, 2) Music, and 3) Babble. 

In each augmentation, one noise file is randomly selected in the MUSAN database and added to the recording with [0, 5, 10, 15] dB SNR.
Alternatively, one music file is randomly selected from the same database and added to the recording with [5, 8, 10, 15] dB SNR. Otherwise, one human babble speech is selected and added to the recording with [13, 15, 17, 20] dB SNR. We repeat the \textit{progressive} data augmentation process with a different random seed to make each copy of the augmentation contain about 2M chunks.

\noindent\textbf{SpecAug}. SpecAug is also investigated after feature extraction~\cite{park2019specaugment}. We apply time warping and time frequency masking at the feature level. The time warping window length is 10. The maximum time mask width is 5 and the maximum frequency mask is 3. Each time we apply two randomly selected time masks and two frequency masks on the features to generate two augmentations.

\vspace{-1ex}
\subsection{Model configuration and training}
The encoding model is a standard TDNN~\cite{snyder2018x} model. 
It has five layers of 1-D CNN and uses a statistical pooling layer to encode the variable-length feature input to a fixed 3000 dimension vector. 
The MLP projection head contains two 512-d layers. After the first layer, BatchNorm and ReLU is applied. L2 norm is applied at the last hidden layer to normalize the embedding to the unit length.
The model is trained on the VoxCeleb2 \textit{dev} split for 150 epochs with a batch size of 4096.
We use the SGD optimizer with 0.9 momentum and initialize the learning rate as $10^{-1}$. The learning rate is reduced to $10^{-4}$ with a cosine learning rate scheduler. The extracted utterance-level embedding size is 512. A dot-product scoring function is applied to compute the similarity between two embeddings. 

For ProtoNCE, we pre-train the model with a MoCo InfoNCE loss for the first 60 epochs. After the 60$^{th}$ epoch, clustering is performed on the whole set using the FAISS~\cite{JDH17} GPU accelerated k-means algorithm. The number of classes $M$ is set to 5000. The final performance is not very sensitive to the cluster number. For each sample $\vct{v_i}$, we choose the same class prototype as the positive, and sample $R=10000$ negative prototypes with replacement from the rest disjoint prototypes. We jointly train the model for another 90 epochs with the loss $\mathcal{L}_{joint}$ in Eq.(\ref{protonce}). For the semi-supervised setting, $\lambda$ is set to $9$ in our experiments. In each mini-batch, 10\% of samples are labeled, and the others are unlabeled.


\section{Results and Discussions}
\label{sec:result}

\subsection{Unlabeled condition}

To thoroughly evaluate our proposed self-supervised methods, we conduct experiments and describe the results and analysis in this section. In \cref{tab:sv_self}, we show the speaker verification evaluation results on the Voxceleb1 test set. Cosine Distance Scoring (CDS) is applied to evaluate the performance.
    \vspace{-2ex}
  
  \begin{table}[htbp]
    \centering
     \caption{Self-supervised SV results on the VoxCeleb1 test set using proposed contrastive methods.}
        \vspace{-2ex}
     \begin{adjustbox}{max width=1\linewidth}

     \begin{tabular}{l|ccc}
          \toprule
         Model   & \multicolumn{1}{c}{CDS EER (\%)} & \multicolumn{1}{c}{DCF}   \\
      \hline

      Disent~\cite{nagrani2020disentangled} & 22.09    & N/A  \\
      CDDL~\cite{chung2020seeing} & 17.52     & N/A   \\
      GCL~\cite{inoue2020semi} & 15.26      & N/A   \\
      i-vector~\cite{huh2020augmentation} & 15.28 & 0.63 (p=0.05)   \\
      AP~\cite{huh2020augmentation} & 9.56     & 0.51 (p=0.05)   \\
      AP+AAT~\cite{huh2020augmentation} & 8.65     & 0.45 (p=0.05)   \\
      \hline
      SimCLR  & 18.14 & 0.81   \\
      MoCo & 15.11 & 0.80   \\
      MoCo + SpecAug & 15.50 & 0.81   \\
      MoCo + WavAug  & \bf{8.63} &  \bf{0.64}  \\
      MoCo + WavAug (ProtoNCE)  & \bf{8.23} & \bf{0.59} \\
      \bottomrule
  
      \end{tabular}%
    \end{adjustbox}

    \label{tab:sv_self}%
  \end{table}%
      \vspace{-2ex}
Methods~\cite{inoue2020semi, huh2020augmentation, nagrani2020disentangled, chung2020seeing} are some recent unsupervised neural methods for speaker representation learning.
 As shown in the table, we can obtain a $15.28\%$ EER using the i-Vector method and $15.11\%$ EER using the MoCo speaker embedding system. It indicates the potential effectiveness of contrastive speaker embeddings.
Moreover, MoCo achieves better results than SimCLR with a large memory queue. There is a relative $16.70\%$ reduction in EER. It meets our assumption that a large amount of negative samples is beneficial to self-supervised contrastive learning. 
Consistent embeddings in the queue with a momentum update mechanism are also essential to the MoCo method.

To investigate the effects of data augmentation, we first compare the MoCo method with a WavAug and a SpecAug module, respectively. We find that our SpecAug module does not help and make the verification results worse. It can be removed in our experiments. The SpecAug module is like a dropout operation on the feature level to improve the model generalization. The Voxceleb test data may not contain lots of variabilities in the spectral domain.
On the contrary, with our progressive WavAug module, the results are improved by a large margin. There is a relative $42.89\%$ improvement in CDS EER from MoCo to the MoCo with WavAug approach. It shows the importance of reducing extrinsic information, such as channel and noise characteristics.
With ProtoNCE, there is another relative $4.63\%$ improvement in CDS EER compared with WavAug. Though the benefit is limited due to the inherent difficulty of speaker clustering on a large dataset, the ProtoNCE loss improves the MoCo training process by alleviating the class collision problem. The EER is also relatively $4.86\%$ lower than the current best reported AP+AAT method without an adversarial augmentation module.

\vspace{-2ex}
\subsection{Visualization of speaker embeddings}
\vspace{-1ex}
\begin{figure}[htbp]
    \begin{minipage}[b]{0.3\linewidth}
        \centering
        \centerline{\includegraphics[height=2.5cm, width=2.9cm]{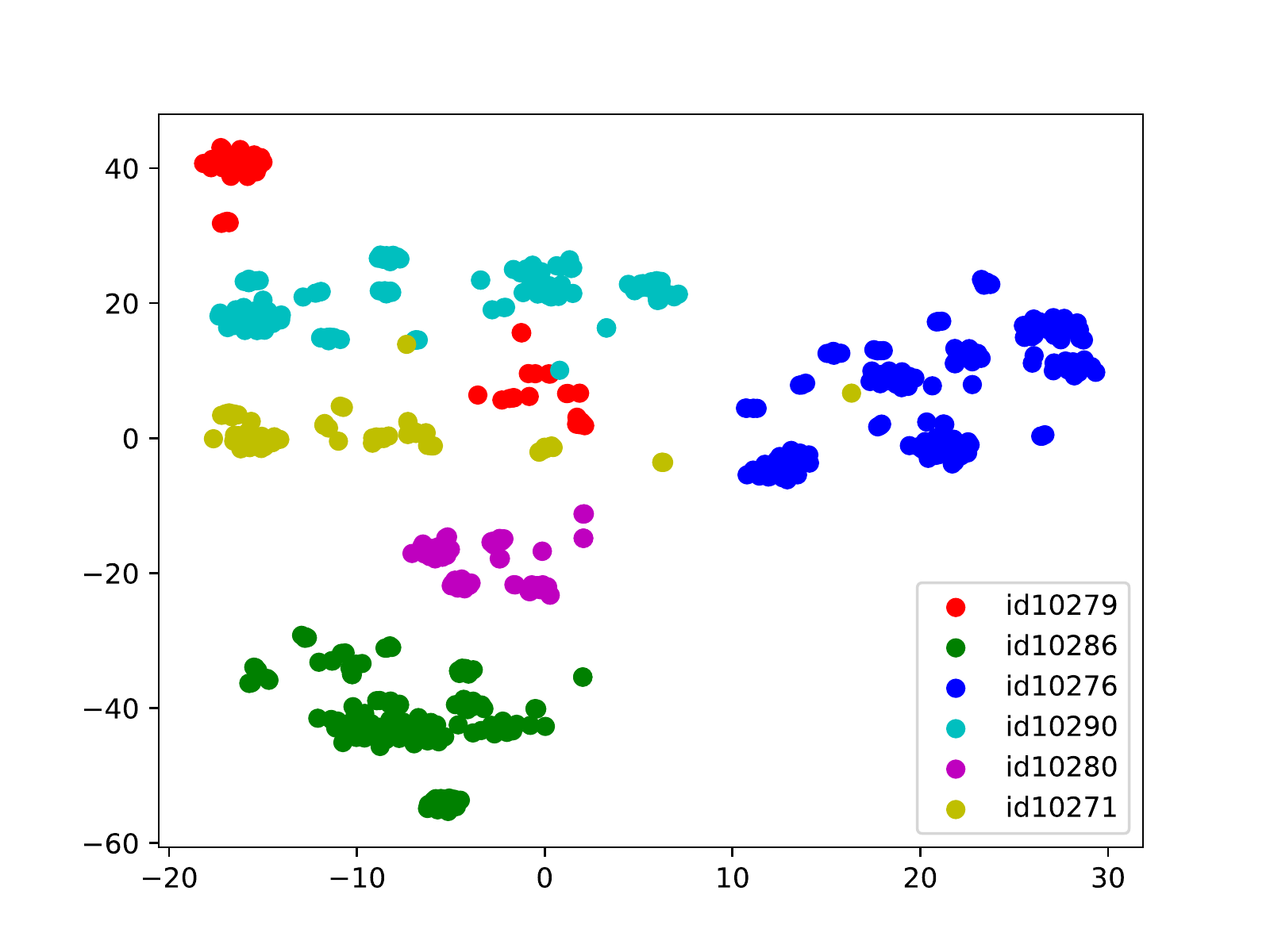}}
        \centerline{\small(a) MoCo }\medskip
      \end{minipage}
      \hfill
      \begin{minipage}[b]{0.3\linewidth}
        \centering
        \centerline{\includegraphics[height=2.5cm, width=2.9cm]{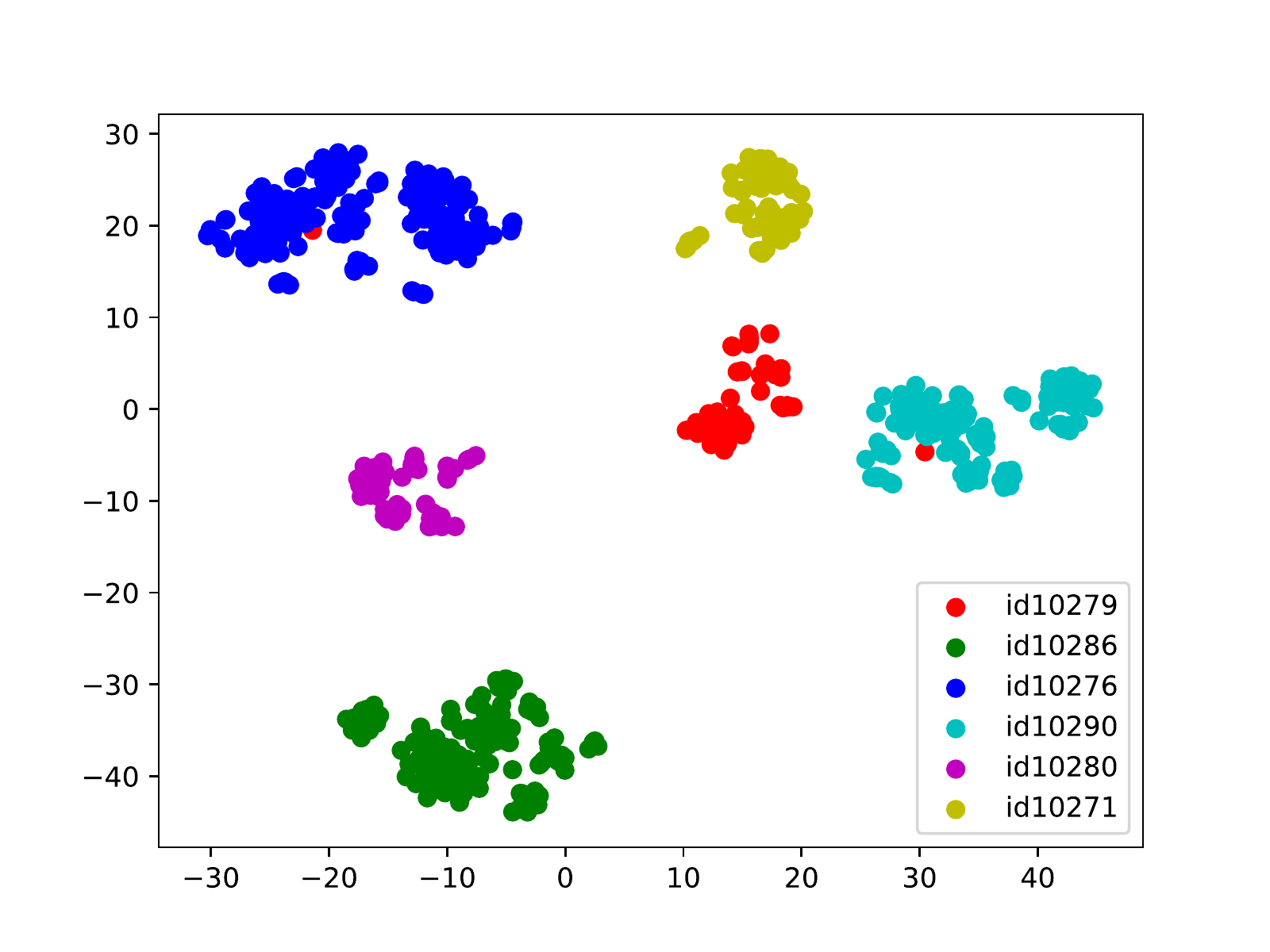}}
        \centerline{(b) WavAug+ProtoNCE}\medskip
      \end{minipage}
      \hfill
    \begin{minipage}[b]{0.3\linewidth}
      \centering
      \centerline{\includegraphics[height=2.5cm, width=2.9cm]{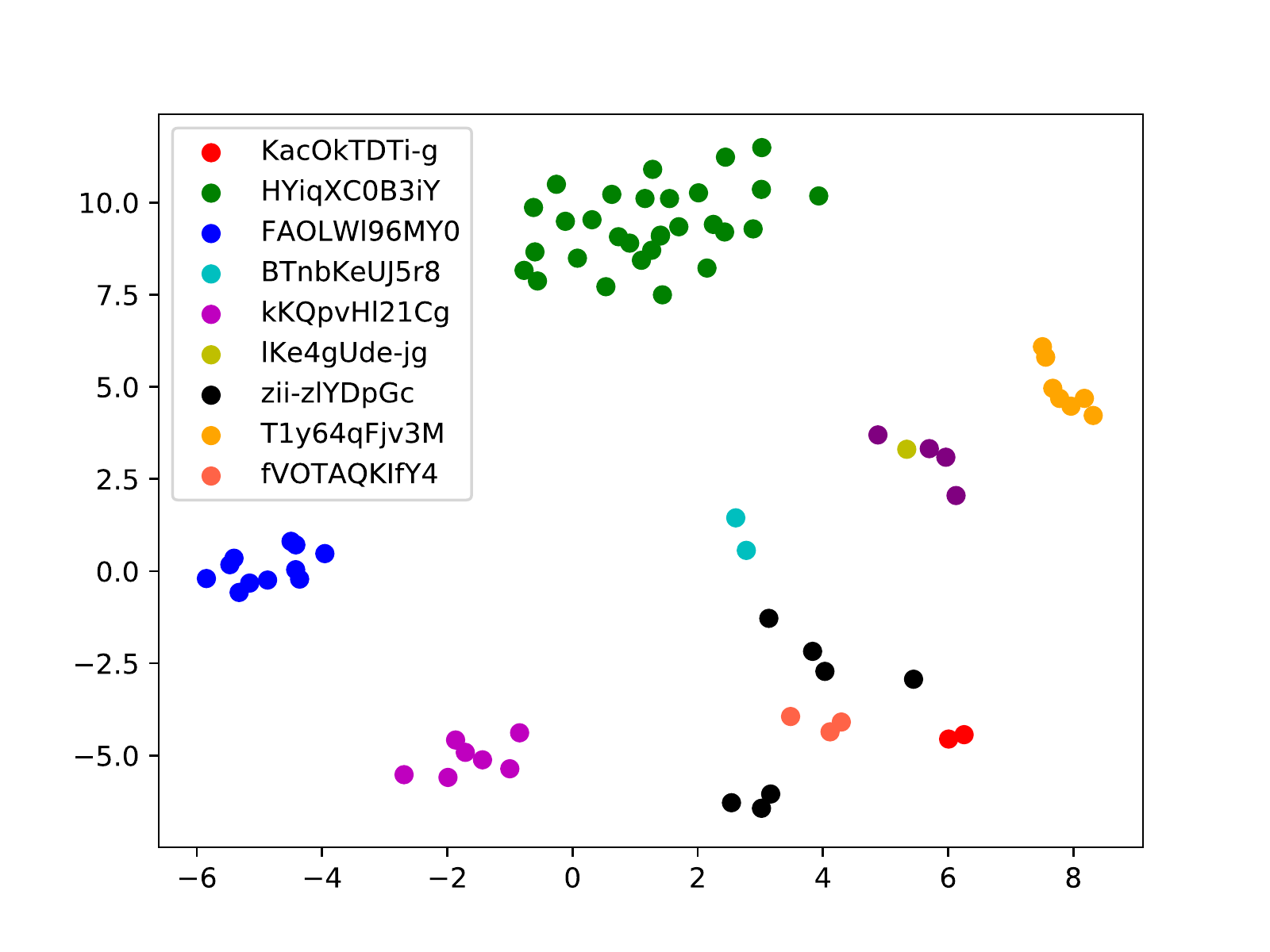}}
      \centerline{(c) Spk-id10290 }\medskip
    \end{minipage}
    \vspace{-2ex}
    \caption{Visualizations of learned speaker embeddings using t-SNE}
    \label{fig:tsne}
    \vspace{-4ex}
\end{figure}
In \cref{fig:tsne}, we visualize the learned self-supervised speaker embeddings of 6 random speakers in the Voxceleb test set.
It can be observed that MoCo speaker embeddings without data augmentation (Fig. 3a) do not distribute well in the embedding space. Utterances from different speakers are grouped together and not well separated. 

MoCo+WavAug+ProtoNCE embedding (Fig. 3b) has much better intra-speaker compactness and inter-speaker separability, which is essential for the speaker verification task. It meets our assumption that WavAug can help significantly reduce channel and noise variabilities.
Also, the test data contains each recording's session id,  reflecting the acoustic condition to some extent. We pick one speaker, ``id10290", from the randomly selected speakers and visualize the embeddings belonging to 10 different sessions. We find that different sessions of the same speaker still tend to be separated. It suggests that there might still be plenty of room to improve the performance if we can further normalize extrinsic variabilities.

\vspace{-2ex}
\subsection{Partially labelled condition}
\vspace{-0.5ex}
We then investigate the performance when there is only limited amount of labeled data. Using the same setting in papers~\cite{inoue2020semi, stafylakis2019self}, we randomly sampled $899$ speakers (15\%) from the Voxceleb2 \textit{dev} set as labeled data. The rest data are used as unlabeled.
  \vspace{-1ex}
\begin{table}[htbp]
    \centering
     \caption{Semi-supervised SV results on the VoxCeleb1 test set. Data augmentations are applied to all approaches.}
        \vspace{-1.5ex}
     \begin{adjustbox}{max width=1\linewidth}
     \begin{tabular}{l|cccc}
          \toprule
  
         Model   & \multicolumn{1}{c}{CDS EER (\%)} & \multicolumn{1}{c}{DCF}  & \multicolumn{1}{c}{PLDA EER (\%)} & \multicolumn{1}{c}{DCF} \\
      \hline
      x-vector (100\% label) & 6.69 & 0.55  & 3.48 & 0.37 \\      
      899-spk x-vector (15\% label)  & 7.91 & 0.70 & 6.39 & 0.60  \\
      MoCo + 899-spk PLDA & 8.63 & 0.64 & 5.21 & 0.49 \\
      \hline
      GCL~\cite{inoue2020semi} & N/A & N/A & 6.01 & N/A \\
      SSL embedding~\cite{stafylakis2019self} & N/A & N/A & 6.31 & 0.53 \\
      SupCon + MoCo (15\% label) & 6.85 & 0.55 & 4.34 & 0.44 \\
      \bottomrule
      \end{tabular}%
    \end{adjustbox}
    \label{tab:sv_semi}%
    \vspace{-0.5ex}
\end{table}%
  \vspace{-2ex}
  
From \cref{tab:sv_semi}, we observe a relative $30.34\%$ improvement from a supervised x-vector trained on $899$ labeled speakers to our proposed semi-supervised \textit{SupCon+MoCo} approach. It shows that with an extra-large amount of unlabeled data, our proposed semi-supervised approach can achieve a better result than a supervised approach trained on partially labeled data.
Moreover, compared with a self-supervised MoCo model trained on all data with a 899-speaker PLDA, our proposed semi-supervised approach obtains a relative $16.70\%$ improvement in EER. In the partially labeled condition, we outperform the semi-supervised GCL and SSL embedding and can approach fully supervised results of the whole labeled set.

\vspace{-1ex}
\section{Conclusions}
\label{sec:conclusion}
\vspace{-1ex}
In this study, we introduced a momentum contrastive speaker representation framework with carefully designed augmentation strategies for text independent speaker verification. To alleviate the class collision problem in self-supervised contrastive learning, we further used clustering to obtain a prototype memory bank and formulated the contrastive learning as an iterative self-labeling process.
The proposed methods were evaluated with two augmentation modules. WavAug was particularly crucial in our proposed self-supervised framework. Moreover, we generalized the self-supervised contrastive loss to the semi-supervised condition and obtained further performance gain. Additional analysis in the learned speaker embedding space demonstrated the effectiveness of our proposed method. At the same time, we still observed the existence of session variability, which restricted the performance of current systems and provided a future direction to be exploited for finer analysis.

\vfill\pagebreak


\bibliographystyle{IEEEbib}
\bibliography{icassp21_ssl}


\end{document}